\documentclass[usegraphicx]{mn2e}
\usepackage{times}
\usepackage{calc}

\def\oversim#1#2{\lower0.5pt\vbox{\baselineskip0pt \lineskip-0.5pt
     \ialign{$\mathsurround0pt #1\hfil##\hfil$\crcr#2\crcr\sim\crcr}}}
\def\Msol{\mbox{${M}_\odot$}}
\def\Lsol{\mbox{${L}_\odot$}}

\newcommand{\lpup}{L$_2$~Pup}
\newcommand{\lpuppis}{L$_2$~Puppis}

\title[Solar-like oscillations in L$_2$~Pup]
{The light curve of the semiregular variable L$\bf_2$~Puppis: II. Evidence for
solar-like excitation of the oscillations}
\author[T.R. Bedding et al.]
       {T.R.~Bedding,$^1$\thanks{E-mail: \tt bedding@physics.usyd.edu.au}
        L.L.~Kiss,$^1$\thanks{On leave from University of Szeged, Hungary}
	H.~Kjeldsen$^2$
	B.J.~Brewer,$^1$
	Z.E.~Dind,$^1$
	S.D.~Kawaler,$^3$
         \newauthor and 
       	A.A.~Zijlstra$^4$\\
	$^1$School of Physics, University of Sydney 2006, Australia\\
	$^2$Theoretical Astrophysics Center, University of Aarhus,
		DK-8000 Aarhus C, Denmark \\
	$^3$Department of Physics and Astronomy, Iowa State University,
	Ames, IA 50011 \\
	$^4$UMIST, Department of Physics, P.O. Box 88, Manchester M60 1QD, UK
}

\begin{document}

\maketitle

\begin{abstract} 

We analyse visual observations of the pulsations of the red giant  variable
\lpup. The data cover 77 years between 1927 and 2005, thus providing an
extensive empirical base for characterizing properties of the oscillations. The
power spectrum of the light curve shows a single mode resolved into multiple
peaks under a narrow envelope. We argue that this results from stochastic
excitation, as seen in solar oscillations.  The random fluctuations in phase
also support this idea. A comparison with X~Cam, a true Mira star with the same
pulsation period, and W~Cyg, a true semiregular star, illustrates the basic
differences in phase behaviours. The Mira shows very stable phase, consistent
with excitation by the $\kappa$-mechanism, whereas W~Cyg shows large phase
fluctuations that imply stochastic excitation. We find \lpup{} to be
intermediate, implying that both mechanisms play a role in its pulsation. 
Finally, we also checked the presence of low-dimensional chaos and could safely
exclude it. 

\end{abstract}


\section{Introduction}

\lpuppis{} (HR 2748; HIP 34922) is a bright nearby red giant with a
pulsation period of about 140\,d.  In Paper~I (Bedding et al. 2002), we showed
that the star has undergone a remarkable change in mean visual magnitude
over the past century, and is currently undergoing a dramatic decline.  We
argued that the most likely cause is the recent formation of dust in an
extended atmosphere.  In this paper, we discuss the pulsation behaviour of
the star and suggest that convection makes a significant contribution to
the driving mechanism.

The mechanism by which pulsations are excited in long-period variables is a
subject of current interest.  Mira variables have large amplitudes and are very
regular, reflecting the nature of the driving process, which is self-excitation
via opacity variations (i.e. the $\kappa$-mechanism). Semiregular variables
(SRs), by contrast, have smaller amplitudes and less  regular light curves. The
nature of these irregularities is far from being understood. Multiple
periodicity has been found in many cases and is generally interpreted in terms
of multimode pulsations (Kiss et al. 1999, Percy et al. 2001). This
interpretation was also supported by the multiple period-luminosity relations
of SRs in the Magellanic Clouds (Wood et al. 1999, Wood 2000) and by observed
mode changes in several semiregular stars  (Cadmus et al. 1991, Percy \&
Desjardnis 1996, Bedding et al. 1998, Kiss et al. 2000). However, even the
clearest examples of multiply periodic SRs exhibit seemingly irregular
additional changes in the light curves (Lebzelter \& Kiss 2001, Kerschbaum et
al. 2001), so that multimode pulsation with stationary frequency and amplitude
content does not fully explain the light variations in SRs. Other explanations
suggested in the literature include chaotic phenomena (Icke et al. 1992,
Buchler et al. 2004), coupling of rotation and pulsation (Barnbaum et al. 1995,
Soszynski et al. 2004) and dust-shell dynamics (H\"ofner et al. 1995, 2003).
Studies of radial velocity variability in SRs (Lebzelter et al. 2000, Lebzelter
\& Hinkle 2002)  suggested that light variations are dominated by stellar
pulsations, which implies that the observed irregularities must be largely
related to stochastic behaviour of the oscillations.

In semiregulars, it seems plausible that there is a substantial
contribution to the excitation and damping from convection. Indeed,
Christensen-Dalsgaard et al. (2001) have suggested that the amplitude
variability seen in these stars is consistent with the pulsations being
solar-like, i.e., stochastically excited by convection (see Bouchy \&
Carrier 2003 and Bedding \& Kjeldsen 2003 for recent reviews of solar-like
oscillations).
Note that we are following convention by referring to oscillations as
  `solar-like' if, like those in the Sun, they are stochastically excited
  and damped.  This is in contrast to oscillations in ``classical''
  pulsating stars, such as Miras and Cepheids, in which the excitation
  occurs via a feedback in opacity variations (the $\kappa$ mechanism).  We
  stress that in other ways, the oscillations in red giants are definitely
  not solar-like. For example, the Sun oscillates in many modes, both
  radial and non-radial, whereas red giants oscillate in only a handful of
  modes that are thought to be exclusively radial.
In K giants, the subject of Mira-like versus solar-like excitation of the
oscillations has been discussed by Dziembowski et al. (2001). They
attempted theoretical modelling of the oscillations in $\alpha$~UMa and
other red giants and suggested that there might even be a mixture of
Mira-like and solar-like modes in certain cases.  The question of
solar-like oscillations in SRs was discussed by Bedding (2003), who
interpreted the Lorentzian shape of the power spectra of several bright SRs
as evidence for stochastically excited oscillations. Here we further
develop this idea by analysing the decades-long visual light curve of
\lpup.

\section{Data analysis}

\begin{figure}
\includegraphics[width=\linewidth]{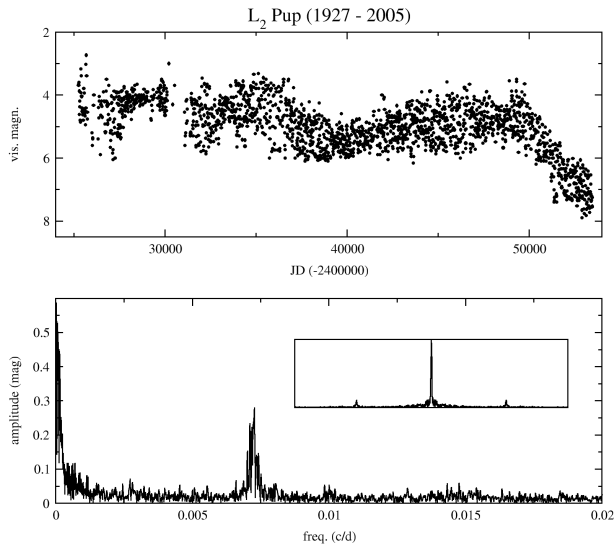}
\caption{The full light curve of \lpup, showing the gradual mean brightness
variations as 10-day means (top) and the corresponding amplitude spectrum 
(bottom). The small
insert shows the spectral window (in amplitude), with sidelobes at $\pm$1 cycle
per year (0.0027 c/d) reflecting annual gaps in the data.}
\label{l2-all}
\end{figure}

The updated light curve of \lpup{} is plotted in the top panel of Fig.\
\ref{l2-all}. The observations since JD 2451000 were made by A.J. Jones, P.
Williams and L.L. Kiss. The dataset consists of 1981 points between
JD 2425249 and 2453487, each one being a 10-day average. This is
roughly 1000 days longer than in fig.\ 1 of Paper~I and it clearly shows
that the latest dimming event is still in progress, though it may have
slowed down a little bit. The frequency spectrum (bottom panel in Fig.\
\ref{l2-all}) is dominated by a set of low-frequency peaks, caused by the
slow variations of the mean brightness, and the main group of closely
separated peaks at the pulsation frequency $\sim$0.007 c/d. There is also a
slight indication of an increased amplitude density around $\sim$0.015 c/d,
which may be attributed to either a harmonic of the main pulsation
(indicating non-sinusoidal variations) or a second (overtone) pulsation
mode.

\begin{figure}
\includegraphics[width=\linewidth]{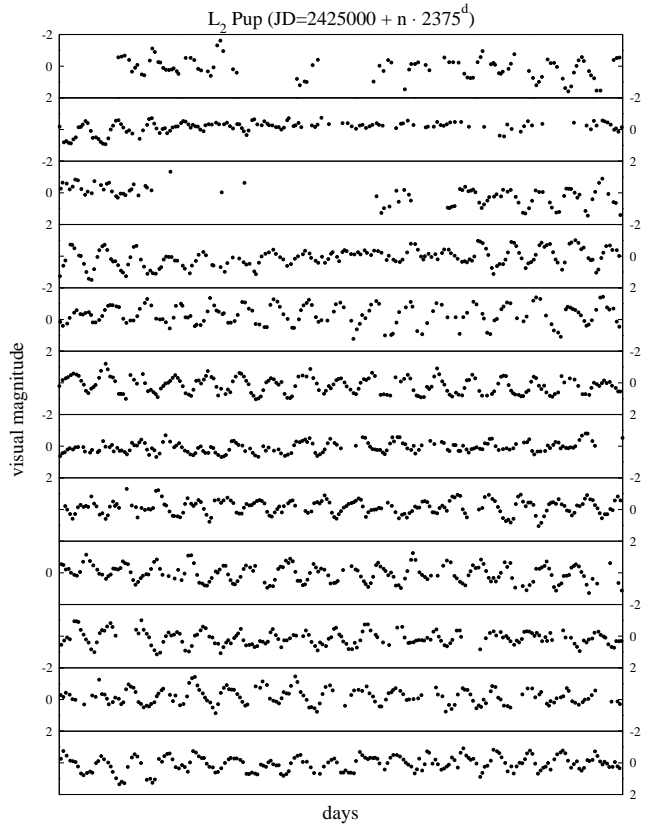}
\caption{Oscillations in \lpup{} (after removal of the long-term trend). 
Each segment 
shows a 2375-d long subset,
which covers roughly 17 cycles. The mean photometric uncertainty is 
about the same as the symbol size.}
\label{l2-panels}
\end{figure}

Details of the light curve can be seen in Fig.\ \ref{l2-panels}. Here, we
subtracted the mean trend of the light curve by fitting a low-order polynomial
to the data. Individual cycles have peak-to-peak amplitudes ranging from
practically zero up to over 2 magnitudes, with no cycles having the same shape.
This illustrates very well the semiregular nature of \lpup. There seems to be
no correlation between the actual mean brightness and amplitude or cycle shape,
which supports the idea that gradual brightness changes are not coupled to the
pulsations.

The data were submitted to the following analysis:

\begin{itemize}
\item[{\it (i)}] the mode lifetime was determined from a power spectrum fit;

\item[{\it (ii)}] phase variations were compared to a true Mira star with 
the same period;

\item[{\it (iii)}] the presence of low-dimensional chaos was tested with
phase space reconstruction. 
\end{itemize}

\section{Discussion}

\subsection{Power spectrum and mode lifetime}

\begin{figure}
\includegraphics[width=\linewidth]{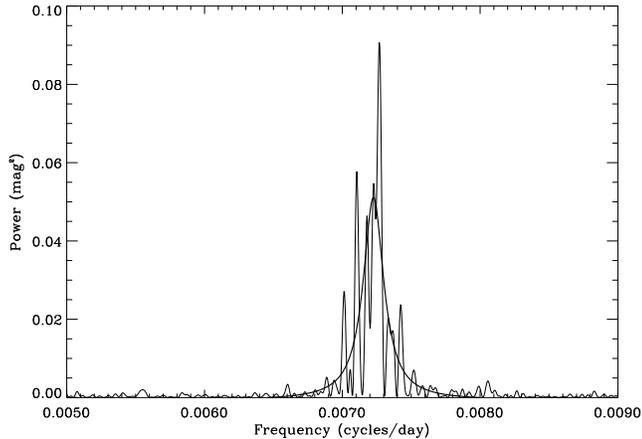}
\caption{\label{fig.power} Close-up view of the power spectrum around
the frequency of pulsations.}
\end{figure}

The power spectrum of the light curve is shown in  Fig.~\ref{fig.power}. We
show a close-up of the region corresponding to the 140-day pulsation period,
and see that the power is split into a series of peaks under a narrow
envelope.  Note that the long term trends in the light curve, which we discussed
in detail in Paper~I, are irrelevant here because they occur at much lower
temporal frequencies.  We have verified that removing these long-term
variations by subtracting a low-order polynomial has neglible effect on the
power spectrum in Fig.~\ref{fig.power}.

The power distribution in Fig.~\ref{fig.power} is typical of a stochastically
excited oscillator and is strikingly similar to close-up views of individual
peaks in the power spectrum of solar oscillations. Assuming a stochastically
excited damped oscillator, we have fitted a Lorentzian profile assuming
$\chi^2$ statistics with two degrees of freedom. This is based on a maximum
likelihood fit assuming an exponential distribution of the noise (Anderson et
al. 1990, Toutain \& Fr\"ohlich 1992).  The fit gives a period of 138.3\,d with
a half-width at half power of $\Gamma= 9.0\times10^{-5}$\,d$^{-1}$, which
implies a mode lifetime of $\tau=(2\pi\Gamma)^{-1} = 4.8$\,yr (i.e.
approximately 12.5 pulsation cycles).  

This represents one of the first measurements of mode lifetime (or damping
time) in a star with solar-like oscillations.  Apart from the Sun itself, in
which  the lifetime is a 2--4 days (e.g. Chaplin et al. 1997), there are
recent  measurement of oscillation lifetimes in four stars.  Bedding et al.
(2004) and Kjeldsen et al. (2005) reported mode lifetimes of 2--3 d in both
$\alpha$~Cen A and B.  For the G giant $\xi$~Hya Stello et al. (2004) suggested
a lifetime of $\sim$2 days (about 20 mean cycle lengths), which is  much
shorter than  was calculated theoretically (Houdek \& Gough 2002). Finally, in
the K giant  Arcturus, which is  the closest example to \lpup, photometric data
taken by the star tracker on the WIRE satellite implied strongly damped
oscillations with a lifetime comparable to the mean period of 2.8 days (Retter
et al. 2003).  Since the damping rate depends on the stellar structure and
convection properties in a complex way with a number of weakly constrained
theoretical parameters (e.g. Balmforth 1992), and moreover, there are no
theoretical calculations directly applicable to \lpup, it is currently
impossible to qualify the agreement with  theoretical expectations. We will
address the dependence of mode lifetime  on main physical parameters in a
subsequent study of bright semiregular variables.

\subsection{Oscillation amplitude and phase}

Could the oscillations in \lpup{} be driven entirely by convection, without any
input from the $\kappa$-mechanism?  This has been suggested for semiregular
variables by Christensen-Dalsgaard et al. (2001), who described such
oscillations as solar-like.  The expected amplitude is highly uncertain, but we
may extrapolate from solar-type stars using empirical scaling laws. Visual
photometric amplitudes of red giant stars are highly dependent on the mean
temperature, whereas velocity measurements, when available, give a more direct
measure of the physical pulsation amplitude. Kjeldsen \& Bedding (1995)
proposed that the velocity amplitude of solar-like oscillations should scale as
$L/M$.  More recently, Kjeldsen \& Bedding (2001) suggested a revised scaling
relation (to account for low observed amplitudes in F stars), in which velocity
amplitude scales as $1/g$ (which is proportional to $L/(M T_{\rm eff}^4)$).

Adopting $L=1500\,\Lsol$, $T_{\rm eff} = 3400$\,K and $M=1\,\Msol$ (Jura et al.
2002), the original and revised scaling relations predict velocity semi-amplitudes
for \lpup{} of about 400\,m/s and 3\,km~s$^{-1}$, respectively. From the
observational side, there are two published radial velocity measurements  for
\lpup: Cummings et al. (1999) measured a semi-amplitude of about
2.5\,km~s$^{-1}$, while Lebzelter et al. (2005) reported six  data points in a
full range of 12\,km~s$^{-1}$ over almost three pulsation cycles, implying a
semi-amplitude of 6\,km~s$^{-1}$. Given the large uncertainties of the
extrapolated relations, the observed amplitudes  do not contradict predictions.
It therefore seems possible that the oscillations in \lpup{} are excited
entirely by convection, with no Mira-like contribution. 

\begin{figure}
\includegraphics[width=\linewidth]{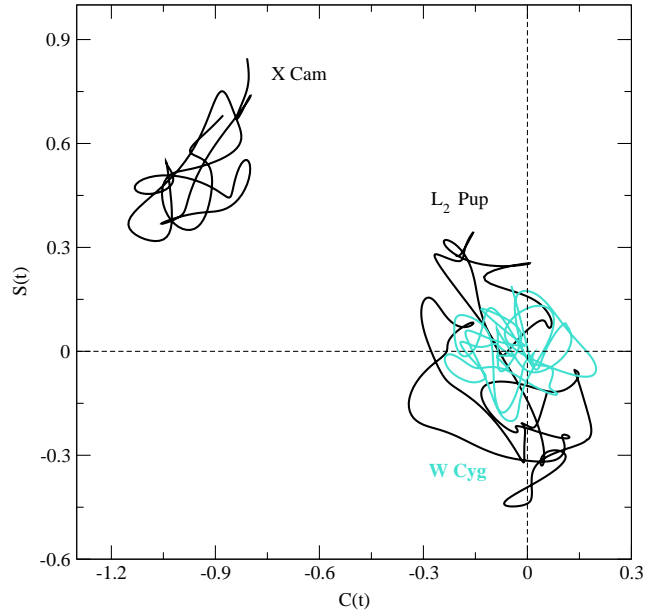}
\caption{Phase variations in \lpup, W~Cyg and X~Cam. $C(t)$ and $S(t)$ are the components
of the G\'abor transform, which give the instantaneous phase as 
$\varphi(t)=\arctan(S(t)/C(t))$.}
\label{l2phase}
\end{figure}

We find further evidence for non-Mira behaviour from a comparison of phase
changes in \lpup{} with those of a true Mira with a very similar pulsation
period.  A well-observed northern circumpolar star is X~Cam, for which the
General  Catalogue of Variable Stars (Kholopov et al. 1985-1988) lists
$m_V=7\fm4-14\fm2$, $P$=143.56\,d. We downloaded its visual data collected by
the Association Fran\c caise des Observateurs d'Etoiles Variables 
(AFOEV\footnote{ftp://cdsarc.u-strasbg.fr/pub/afoev/}), extending back to the 
early 20th century, and binned them into 10-day averages. Also, as an
example of a ``real'' semiregular, we took visual data of W~Cyg, an SRb
type star with two dominant periodicities, 131\,d and 235\,d (Howarth 1991).
Then we calculated the following quantities for each star:

\begin{equation}
C(t) = \sum_{i=1}^{n} w(t,t_i) (m(t_i)-\langle m(t_i) \rangle) \cos(2 \pi f t_i)
\end{equation} 

\begin{equation}
S(t) = \sum_{i=1}^{n} w(t,t_i) (m(t_i)-\langle m(t_i) \rangle) \sin(2 \pi f t_i)
\end{equation} 

\noindent which are closely related to the Fourier transform of the light curve
$\{m(t_i)\}$ $(i=1...n)$. The only difference is the presence of the Gaussian
weight-function $w(t,t_i)$, which was  used as a moving window over the light
curve (in other words, $C(t)$ and $S(t)$ are the real and the imaginary 
components of the G\'abor transform; for a recent review of related transforms 
see Buchler \& Koll\'ath 2001). In our case, $w(t,t_i)$ was centered at $t$ and
ran between $t_1$ and $t_n$ with a time step of 50 days, and the full-width at 
half maximum was fixed as 700 days (about 5 cycles). This way we could measure
the local phase $\varphi(t)=\arctan (S(t)/C(t))$ at fixed frequency $f=1/P$.
For X~Cam we determined $P$=143.69\,d, very close  to the GCVS value, while for
\lpup{} we used the mean period given by the center of the fitted Lorentzian
(138.3\,d). In the case of W~Cyg, we took the shorter period (131\,d), which is
the dominant one. As usual, $\langle m(t_i) \rangle$ denotes the mean
magnitude. 

\begin{figure}
\includegraphics[width=\linewidth]{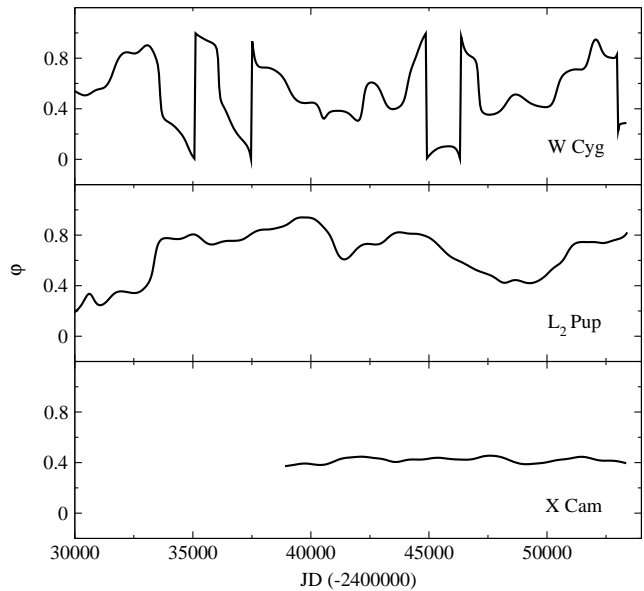}
\caption{Phase variations in \lpup, W~Cyg and X~Cam as function of time.}
\label{phases}
\end{figure}

First we show the phase variations by plotting $S(t)$ versus $C(t)$ in Fig.\
\ref{l2phase}.  The fact that \lpup{} and W~Cyg are closer to the origin than
X~Cam is a simple consequence of their smaller light curve amplitudes
(proportional to $\sqrt{C(t)^2+S(t)^2}$). The interesting thing is the phase
range covered by the stars. While X~Cam was meandering over a fairly narrow
range of about 20--30$^\circ$, drawing a loosely defined arc in the phase
plane, \lpup{} went almost all around the full circle. We interpret this
difference as evidence that in \lpup{}, the changes in both amplitude and phase
are close to random, almost as randomly as in W~Cyg, which is 
centered on the origin.

The intermediate nature of \lpup{} is also apparent from the phase variations 
as functions of time (Fig.\ \ref{phases}). X~Cam's phase is very stable in time,
with no sudden changes; \lpup{} has good phase stability over several thousand
days but then some jumps occur. In contrast, W~Cyg shows
random phase variations on short time scales (see Howarth 1991 for further 
discussion of the phase changes in this star). 
The phase variations of X~Cam are fully consistent with the $\kappa$-mechanism,
which by definition includes a phase coherent positive feedback from periodic
opacity changes. The phase of W~Cyg varies continuously, supporting the view
that the oscillations are stochastically excited, presumably by convection.
Finally, \lpup is an intermediate case. The phase fluctuations imply
stochastic behaviour, but there may also be some driving from the
$\kappa$-mechanism.

\subsection{A test for low-dimensional chaos}

Could the seemingly complex pulsational behaviour of \lpup{} be caused by a
simple low-dimensional chaotic system?  Recently, Buchler et al. (2004) 
concluded that the irregular
pulsations in some semiregulars are the result of the non-linear
interaction of two strongly nonadiabatic pulsation modes, although they
admitted that stochastic procedures may also affect the light curve shape
to some extent.  The first clear detection of chaos in a Mira star was
presented by Kiss \& Szatm\'ary (2002), who demonstrated the chaotic origin of the
amplitude modulation of R~Cygni.  Here we employ the same non-linear tools
of light curve analysis in order to yield insights into the dynamics of the
pulsations in \lpup.  Instead of using the whole dataset, we restricted the
nonlinear analysis to the better-sampled two-thirds of the data, from
JD~2435000 onwards.

\begin{figure}
\includegraphics[width=\linewidth]{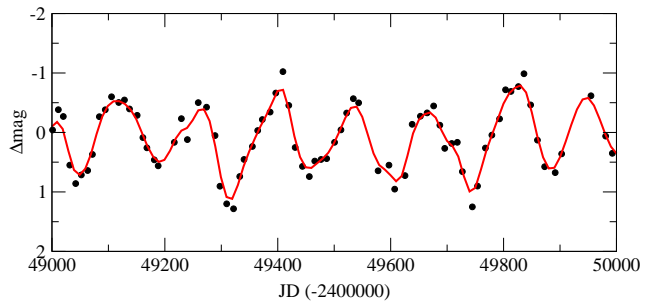}
\caption{\label{fig.polyfit} A typical light curve
segment with the smoothed and interpolated signal.}
\end{figure}

\begin{figure*}
\includegraphics[width=14cm]{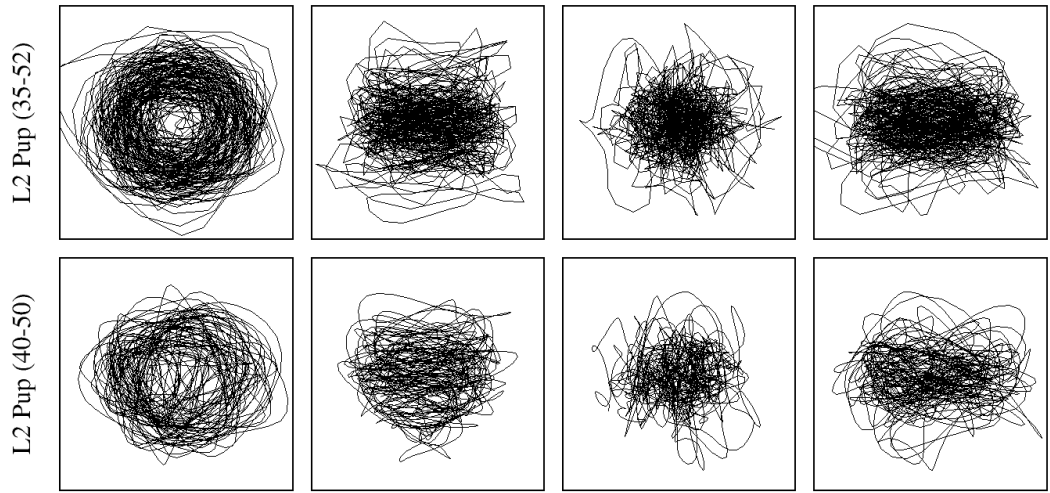}
\caption{\label{fig.bkproj} Broomhead-King projections. See text for further
details.}
\end{figure*}

We made the following pre-processing steps: 1.~long-term trend removal by
subtracting a polynomial that was fitted to the binned light curve;
2.~smoothing and interpolating the resulting light curve.  The first step
was made to remove the long-term mean brightness variations that are caused
by the increased dust extinction events and are not connected with
pulsation (Paper~I).  We experimented with different orders of polynomials
and accepted 10 as the best one which followed quite well the sudden
changes of the mean level.  After
subtracting the polynomial, the residual light curve clearly showed the
complex amplitude changes.
Further interpolating consisted of fitting Akima splines to the data with
the constant sampling rate of 10 days.  The spline fit was smoothed with a
Gaussian weight-function with an FWHM of 20 days. We show a part of the
final result in Fig.~\ref{fig.polyfit}.

The pre-processed light curve was submitted to our nonlinear analysis.  The
applied numerical tools and their basics have been described by
Hegger et al. (1999).  Briefly, we used time-delay embedding to reconstruct the
phase space of the system.  First, we estimated the embedding parameters,
the optimal time-delay embedding dimension.  We then calculated various
projections to visualize the trajectories in the phase space.  The
existence of an attractor would produce regular structures, and finding
these was the aim of the analysis.

As for R~Cygni, the optimal delay was found to be between 10--30\% of the
formal period.  Since the reconstructed images did not show strong
dependence on the time delay, we adopted 50 days, corresponding to 5 steps
in the dataset and about 30\% of the mean period.  We tried different
embedding dimensions ($d_{\rm e}$) and the most informative images were
found for $d_{\rm e}$=4.  To visualize the embedded state vectors, we have
generated the Broomhead-King projections (Broomhead \& King 1986), which project onto
the eigenvectors of the correlation matrix.

The most informative projections are shown in the top row of
Fig.~\ref{fig.bkproj}.  For comparison, we also show the same projections
for a shorter dataset without long-term trend removal but with the same
smoothing and interpolation procedure.  This subset was chosen between JD
2440000 and JD 2450000, when the mean brightness of \lpup{} was fairly
constant.  Both datasets yield the same result: there are no regular
structures in the reconstructed phase space of \lpup{} and the presence of
an attractor can be safely excluded.  The system is temporarily close to
stable states from time to time (as suggested by the circular structures in
the first column of Fig.~\ref{fig.bkproj}), but it then changes erraticly
and the phase space projections become fuzzy clouds.  We have checked this
conclusion up to $d_{\rm e}=10$ and the results did not change, producing
only similar blurred structures in higher dimensions.  We conclude that the
complex light variations cannot be explained by low-dimensional chaos.

We also note that Buchler et al. (2004) made a somewhat extrapolative argument
against stochasticity in semiregular variables. While criticizing Konig et al.
(1999) for interpreting R~Scuti's light curve in terms of stochastic damped
oscillators,  Buchler et al. (2004) rejected the idea as being meaningless on
physical grounds, because no mechanism is proposed that could excite damped
modes to such large amplitudes, like the factor of 40 in the light curve of
R~Sct (i.e. that star's visual amplitude can reach 4 mags). Furthermore, they
also considered the energetics of a 0.7 M$_\odot$ pulsation model with L=1000
L$_\odot$ and T$_{\rm eff}$=5300 K (implying 37 R$_\odot$ for model radius).
Their main argument was the fact that the model's average pulsational kinetic
energies in alternating large and small amplitudes exceeded by a factor of two
the average turbulent energies, so that even if all available turbulent energy
were converted into pulsational kinetic energy, there is not enough energy.
However, while this consideration might be true for R~Sct, which is an RV~Tauri
type pulsating yellow supergiant, both the large amplitude in R~Sct and the
quoted model properties are highly irrelevant when considering pulsations in
red giant stars. Firstly, the visual amplitude of semiregular and Mira stars
are grossly enhanced by the extreme temperature sensitivity of molecular
species in their much cooler atmospheres with typical temperatures around
3,000--3,300 K. In addition, the visual range is in the Wien-part of the
spectrum for these stars, which also increases the amplitude's temperature
sensitivity. For that reasons, infrared observations (e.g. in K band) draw a
more reliable picture of their luminosity fluctuations. Typical near-infrared
amplitudes in semiregular stars are between 0.1-0.3 mag (Whitelock et al. 2000,
Smith 2003), which means there is no need for any exotic mechanism to excite
damped modes to large amplitudes. For \lpup{}, the DIRBE instrument on the COBE
satellite measured less than 10\% fluctuations at  2.2$\mu$ (Jura et al. 2002,
Schmidt 2003), which is typical among Mira-like semiregulars (i.e. SRa type
SRs). Secondly, we feel that the applicability of a 5300 K model to 3000--3300
K stars must be limited and speculative. Red giant stars can have  luminosities
and radii larger by up to a factor of ten, in which parameter region the
properties of convection are not necessarily the same as in hotter and smaller
stars. Consequently, while Buchler et al. (2004) presented convincing evidence
for low-dimensional chaos in three (or maybe four) semiregular stars, their
arguments against stochastic excitation do not lie on firm physical basis. We
therefore believe that  stochastic (solar-like) oscillations offer the best
explanation for the complex variability in \lpup.  

\subsection{Pulsation mode}

Jura et al. (2002) cited the difference between 12\,$\mu$m light curve and
those at $H$ and $K$ for \lpup{} as evidence that the pulsations are
non-radial. Smith (2003) also noted the intriguing difference between the
shorter and longer wavelength infrared observations of the DIRBE instrument,
which was exceptional in a sample of 207 infrared sources. In particular, the
maxima at 1.25\,$\mu$m preceded those at 4.9\,$\mu$m by 10--20 days, while at
4.9\,$\mu$m, there was a secondary peak between the two 1.25\,$\mu$m maxima.
Also, the 2.2\,$\mu$m and 3.5\,$\mu$m light curves resembled the 1.25\,$\mu$m
curve, while the 12\,$\mu$m curve was similar to that at 4.9\,$\mu$m. The Jura
et al. hypothesis was based  on the assumption that, while the 12\,$\mu$m flux
tracks the light emitted by the circumstellar dust, and hence  measures the
luminosity of the entire star, the near-infrared fluxes measure only emission
from the hemisphere of the star that faces the Earth. If there are differences
then the photosphere must vary in time nonspherically, implying non-radial
pulsation, which would also naturally explain the marked time variation of the
position angle of the net polarization of the star (Magalhaes et al. 1986).

However, as we have shown in Paper I, the extinction-corrected $K$ band
magnitude of \lpup{} is very close to the value expected from the Mira P--L
relations. The good agreement tends to argue against the hypothetic nonradial
pulsations (unless one allows that Miras also pulsate non-radially). 
The full velocity amplitude of 12 \,km~s$^{-1}$ (Lebzelter et al. 2005)
is also hardly compatible with non-radial oscillations, which are not expected
to result in  large radial velocity amplitudes (Wood et al. 2004). Looking at
DIRBE light curves of Mira stars, it is apparent that not only are there
similar phase lags in Mira stars (Smith et al. 2002), but also that small
differences do exist between the near- and mid-infrared data. As discussed by
Smith et al. (2002), there are no theoretical models that predict these phase
lags, because of the complexity of the problem. In addition to the
pulsation-induced luminosity variations, the 12\,$\mu$m light curve also
reflects the dynamics of dust formation process which is governed by a time
scale of its own (Winters et al. 2000 and references therein). Furthermore, the
DIRBE observations were taken a few pulsation cycles before the beginning of
the major dimming event in 1994. It seems to be likely that at the time of the
DIRBE observations, there had already been some activity in dust formation,
which can explain the extra features in the mid-infrared data without invoking
non-radial pulsations. For example, Soker (2000) examined the effects of
magnetic cool spots, above which large quantities of dust are expected to form.
Alternatively, interactions with a close binary component (but outside the AGB
envelope) may yield a strong density contrast in the equatorial plane, with
similar outcomes. While  currently there is no evidence for binarity in \lpup,
Ireland et al. (2004), using optical interferomety, found evidence for highly
clumpy structures in the  circumstellar dust shell as close as the dust
condensation radius. Very recently, Wood et al. (2004) presented observations
that suggested large scale star spot activity in some semiregulars with long
secondary periods. Although \lpup{} does not belong to that subclass of SRs, we
think that  the Soker (2000) model offers a more realistic mechanism to produce
inhomogeneous dust and therefore different mid-infrared light curves in \lpup{}
than non-radial pulsations.  

\section{Conclusions}

\lpup{} is a very interesting semiregular star. Its light curve is dominated by
two completely different mechanisms: gradual dimmings caused by circumstellar 
dust and pulsations. These two are completely independent, as we did not 
find any evidence for coupling between them. The present paper discussed
properties of the pulsations, outlining the physical implications. 

The frequency spectrum of the long-term light curve shows a peak  of power that
is resolved into multiple peaks under a narrow envelope. This is  consistent
with stochastic excitation, as seen in solar oscillations. The mode life-time
can be derived from the width of the envelope: for \lpup{} this is found to be
about 5 yr. The oscillation amplitude agrees roughly with the predictions of
simple scaling laws, while the phase behaviour is markedly different from a
Mira star with the same pulsation period. The seemingly random amplitude and
phase changes argue against excitation by the $\kappa$-mechanism alone,
although it might make some contribution.

A test for low-dimensional chaos ruled out the possibility of non-linear 
interactions of a few pulsation modes as the reason for seemingly irregular 
light variations. We also argued against the hypothesis of non-radial pulsations
in \lpup{}: the difference between the near- and mid-infrared data may only be
related to clumpy dust production, possibly driven by spot activity. 

Bedding (2003) has recently given examples of other semiregulars whose power
spectra display similar evidence for solar-like oscillations.  The inferred
mode lifetimes range from a few years, as found in \lpup, down to less than a
year. It seems plausible that stochastic excitation, presumably from
convection, plays an important -- perhaps dominant -- role in the behaviour of
semiregular variables.

\section*{Acknowledgments}

TRB and LLK are grateful to the Australian Research Council for financial 
support. LLK is supported by a University of Sydney Postdoctoral Fellowship.
SDK's participation was partially supported by the U.S. National Science
Foundation through Grant AST-0205983, and the NASA Astrophysics Theory 
Program through grant NRA-98-03-ATP-078.
We are grateful to the hundreds of amateur observers whose
measurements were critical to this paper, particularly Albert Jones and
Peter Williams. We
also made extensive use of the SIMBAD and the ADS data services.

\end{document}